\RequirePackage{ifpdf}
\ifpdf % We are running pdfTeX in pdf mode
\documentclass[pdftex]{sigma}
\else
\documentclass{sigma}
\fi

\begin{document}
\allowdisplaybreaks

\renewcommand{\PaperNumber}{019}

\FirstPageHeading

\ShortArticleName{Nonlocal Symmetries and Generation of Solutions
for PDEs}

\ArticleName{Nonlocal Symmetries and Generation of Solutions\\ for
Partial Dif\/ferential Equations}

\Author{Valentyn TYCHYNIN~$^\dag$, Olga PETROVA~$^\ddag$ and
Olesya TERTYSHNYK~$^\ddag$}

\AuthorNameForHeading{V. Tychynin, O. Petrova and O. Tertyshnyk}

\Address{$^\dag$~Prydniprovs'ka State Academy of Civil Engineering
and Architecture,\\
$\phantom{^\dag}$~24a Chernyshevsky Str., Dnipropetrovsk, 49005
Ukraine}
\EmailD{\href{mailto:tychynin@mail.pgasa.dp.ua}{tychynin@mail.pgasa.dp.ua},
\href{mailto:tychynin@ukr.net}{tychynin@ukr.net}}

\Address{$^\ddag$~Dnipropetrovsk National University, 13 Naukovyi
Per., Dnipropetrovsk,  49050 Ukraine}

\ArticleDates{Received January 06, 2006, in f\/inal form January
17, 2007; Published online February 06, 2007}

 \Abstract{We have constructed new formulae for generation of
 solutions for the nonlinear heat equation and for the Burgers equation
 that are based on linearizing nonlocal transformations and on nonlocal
 symmetries of linear equations. Found nonlocal symmetries and
 formulae of nonlocal nonlinear superposition of solutions of
 these equations were used then for construction of chains of exact solutions.
Linearization by means of the Legendre transformations of a
second-order PDE with three independent variables allowed to
obtain nonlocal superposition formulae for solutions of this
equation, and to generate new solutions from group invariant
solutions of a linear equation.}


\Classification{35A30; 35K55; 35K57; 35L70}

\section{Introduction}
Nonlocal symmetries of nonlinear equations of mathematical physics
stay in the center of attention of many
authors~\cite{anco&bluman,moitsheki&broadbrige&edwards,galas,papachristou&harrison,ibragimov1994}.
Methods used for investigation of dif\/ferential equations (DEs)
include application of a wide spectrum of nonclassical and
nonlocal symmetries of
DEs~\cite{olver&rosenau1986,olver&rosenau1987,
fushchych&serov&tychynin&amerov,krasil'shchik&vinogradov,anderson&ibragimov1979,ibragimov1985}.

Knowing the symmetries we can construct exact solutions for the
equations and then proceed with  their generating, as well as
describe sets of conserved quantities, reduce the initial equation
to the equations with smaller number of variables and solve other
problems related to these equations.

The method of studying of the group invariance properties of the
DEs, created by S.~Lie, has been further developed  by
G.~Birkhof\/f, L.V.~Ovsyannikov, P.J.~Olver, W.I.~Fushchych,
N.H.~Ibragimov, G.~Bluman, J.D.~Cole, P.~Winternitz and many
others.

In the 1970s V.A.~Fok~\cite{fok} found the symmetry of hydrogen
atom in Coulomb f\/ield, which, as became clear later, could not
be found by the Lie method. Then other similar facts started to
arise, and it required an adequate explanation. Resumption of
interest to B\"acklund transformations~\cite{lowner} in the 1950s
and discovery of the Miura transformation allowed explaining of
existence of an inf\/inite set of conservation laws for
Korteweg--de Vries equation, and active period of f\/inding of
further generalizations of S.~Lie theory had started.

One of such generalizations,  conditional symmetries of
DEs~\cite{fushchych&serov}, that are referred the li\-te\-rature
also as nonclassical or weak
symmetries~\cite{olver&rosenau1986,olver&rosenau1987,dzamay&vorob'ev1994},
may be found by adding of some condition to the equation with this
conditional symmetry being the symmetry of the resulting system.
Usually this condition is an equation of a surface invariant under
inf\/initesimal operator admitted by the equation under
investigation~\cite{gandarias&bruzon,clarkson&mansfield}.

Theory of Lie--B\"acklund group transformations was developed by
R.L.~Anderson and N.H.~Ib\-ra\-gimov in the 1970s
\cite{ibragimov1994,anderson&ibragimov1979,ibragimov1985,ibragimov&anderson1977}.
 Such
transformations forming a one-parametrical group depend on
derivatives
\begin{gather}\label{general nonloc.tr-n}
\tilde
x_i=f^{i}(x,u,{\mathop{u}\limits_{1}},\dots,{\mathop{u}\limits_{k}};\varepsilon),
\qquad \tilde
u^a=g^a(x,u,{\mathop{u}\limits_{1}},\dots,{\mathop{u}\limits_{k}};\varepsilon).
\end{gather}
In these formulae and elsewhere $x=(x_0, x_1,\dots, x_n)$ is a set
of independent variables, and $u=(u^1, u^2,\dots, u^m)$ is a set
of dependent variables. The corresponding inf\/initesimal operator
has the following form:
\[
X=\xi^{i}(x,u,{\mathop{u}\limits_{1}},\dots,{\mathop{u}\limits_{k}})
\partial_i+\eta^a(x,u,{\mathop{u}\limits_{1}},\dots,{\mathop{u}\limits_{k}})\partial_{u^a},
\qquad i=0,1,\dots,n, \quad a=1,2,\dots,m.
\]
Lie--B\"acklund symmetries def\/ined in such a way present the
same concept as generalized vector f\/ields in P.J.~Olver's
terminology~\cite{olver1993}. Dependence of expressions
in~(\ref{general nonloc.tr-n}) on derivatives is a reason for
regarding such symmetry as nonlocal. Note that reversion of the
transformation~(\ref{general nonloc.tr-n}) requires integration
procedure.

Further in the paper we  use  the following standard notations:
\begin{gather*}
u_{\mu}=\frac{\partial u}{\partial x^{\mu}}\equiv
\partial_{\mu}u,\qquad  \mu=0,1,2,\dots,n-1,\qquad
\mathop{u}\limits_{1}=\{u_{\mu}\},\nonumber\\
u_{\mu\nu}=\frac{\partial ^{2} u}{\partial x^{\mu}\partial
x^{\nu}}\equiv
\partial_{\mu}\partial_{\nu}u,\qquad  \mu,\nu=0,1,2,\dots,n-1, \qquad
\mathop{u}\limits_{2}=\{u_{\mu\nu}\}.%\label{designations}
\end{gather*}

Another direction in investigation of nonlocal symmetries is based
on representation of the equation
\[
F_1(x,u,{\mathop{u}\limits_{1}},\dots,{\mathop{u}\limits_{k}})=0
\]
in the form of a conservation law
\[
F_i(x,u,{\mathop{u}\limits_{1}},\dots,{\mathop{u}\limits_{k}})=D_1\phi^1(
x,u,{\mathop{u}\limits_{1}},\dots,{\mathop{u}\limits_{r}})+D_2\phi^2(
x,u,{\mathop{u}\limits_{1}},\dots,{\mathop{u}\limits_{r}})=0
\]
 and
on introduction of a new potential variable by the relation
\[
D_2v=\phi^1(x,u,{\mathop{u}\limits_{1}},\dots,{\mathop{u}\limits_{r}}).
\]
It allows studying of a  ``potential symmetry'' of the equation as
a classical symmetry of system~\mbox{\cite{anco&bluman,galas}}
\begin{gather*}
F_i(x,u,{\mathop{u}\limits_{1}},\dots,{\mathop{u}\limits_{k}})=0,\qquad
D_2v=\phi^1(x,u,{\mathop{u}\limits_{1}},\dots,{\mathop{u}\limits_{r}}),\qquad
D_1v=\phi^2(x,u,{\mathop{u}\limits_{1}},\dots,{\mathop{u}\limits_{r}}).
\end{gather*}
It may be noted that in the described method for investigation of
nonlocal symmetries of DE an important place belongs to  the
recursion operator for an equation introduced by
P.J.~Olver~\cite{olver1993}.

Further development of this approach was obtained in~\cite{guthrie
1994, guthrie&hickman 1993} via reversion of the recursion
operator. Nonlocal symmetries of the pseudopotential type in sense
of prolongation structures of Estabrook and Wahlquist were
considered in~\cite{leo&leo&soliani&tempesta}. It was discovered
in the paper~\cite{olver&sanders&wang}  that the Jacobi identity
for characteristics of nonlocal vector f\/ields ``appears to fail
for the usual characteristic computations'' that led to the notion
called ``ghost symmetries''. That also showed impossibility of
discussion of Lie algebraic properties for general nonlocal
symmetries of DEs.

 Utilization of the method of external
dif\/ferential forms by E.~Cartan  allowed K.~Harrison and
C.J.~Papachristou~\cite{papachristou&harrison} to calculate
classical point (internal or geometrical) symmetry by means of
so-called isovectors generated by a vector f\/ield
\[
V=\xi^i(x,u)\partial_i+\eta^a(x,u)\partial_{u^a}.
\]
So the internal symmetry of DE was determined by the requirement
of invariance of an ideal of external dif\/ferential forms of the
system $\gamma_i$ that was generated by the equation under action
of the Lie derivative $\pounds_V$ with respect to vector f\/ield
$V$
\[
\pounds_V\gamma_i=
{b_i}^k\gamma_k+{\Lambda_i}^k\gamma_k+\gamma_k{M_i}^k.
\]

Nonlocal symmetry of the given equation was studied in
\cite{papachristou&harrison} with dependence of factors $V$ on
derivatives included into this scheme.

Geometrical theory of nonlocal symmetries of DEs has been
developed  by A.M.~Vinog\-radov and
I.S.~Krasil'shchik~\cite{krasil'shchik&vinogradov} and formulated
in the language of jet f\/iber bundles for  the functions being
solutions of the given equation.

In this paper we continue investigation of nonlocal symmetries of
PDEs~\cite{fushchych&tychynin,tychynin} by means of nonlocal
transformations of variables
\begin{gather*}
B^p(x,u,{\mathop{u}\limits_{1}},\dots,{\mathop{u}\limits_{k}};y,v,
{\mathop{v}\limits_{1}},\dots,{\mathop{v}\limits_{t}})=0,\\
p=1,2,\dots,n+m,\qquad u=\{u^a\},\qquad a=1,2,\dots,m
\end{gather*}
 or, in a
simpler form
\begin{gather*}
\tau:\left\{\begin{array}{@{}l}
x_i=h^i(y,v,{\mathop{v}\limits_{1}},\dots,{\mathop{v}\limits_{k}}),
\\ u^a=H^a(y,v,{\mathop{v}\limits_{1}},\dots,{\mathop{v}\limits_{k}}).
\end{array}\right.%\label{Z11}
\end{gather*}
Under the transformation $\tau$ the equation
\[
F_1(x,u,{\mathop{u}\limits_{1}},\dots,{\mathop{u}\limits_{n}})=0
\]
is transformed into the new equation
\[
\Phi(y,v,{\mathop{v}\limits_{1}},\dots,{\mathop{v}\limits_{m}})=0
\]
of the order $m=n+k$. Let us present the obtained equation  in the
form of equality
\[
F_1(x,u,{\mathop{u}\limits_{1}},\dots,{\mathop{u}\limits_{n}})=\lambda
F_2(y,v,{\mathop{v}\limits_{1}},\dots,{\mathop{v}\limits_{s}})=0.
\]
Here $\lambda$ is the dif\/ferential operator of the order $r$:
$r+s=n+k$ (so it is possible to carry out factorization of the
equation $
\Phi(y,v,{\mathop{v}\limits_{1}},\dots,{\mathop{v}\limits_{m}})=0
$ by means of $
F_2(y,v,{\mathop{v}\limits_{1}},\dots,{\mathop{v}\limits_{s}})=0$).
In this case we say that the equations
\[
F_1(x,u,{\mathop{u}\limits_{1}},\dots,{\mathop{u}\limits_{n}})=0
\qquad \mbox{and}\qquad
F_2(y,v,{\mathop{v}\limits_{1}},\dots,{\mathop{v}\limits_{s}})=0
\]
are connected by the nonlocal transformation $\tau.$

If invariance algebras of these equations have dif\/ferent
dimensions, we can put ``extra''  symmetries of one equation into
correspondence with nonlocal symmetry of another equation. We use
nonlocal transformations of variables $\tau$ for construction of
formulae for generation of solutions both in the case of nonlocal
invariance of the equation
\[
F_1(x,u,{\mathop{u}\limits_{1}},\dots,{\mathop{u}\limits_{n}})=
F_1(y,v,{\mathop{v}\limits_{1}},\dots,{\mathop{v}\limits_{n}})=0
\]
and for other equations. If the equation
\[
F_2(y,v,{\mathop{v}\limits_{1}},\dots,{\mathop{v}\limits_{s}})=0
\]
is linear and homogeneous, the transformation $\tau$ allows
construction of nonlinear nonlocal superposition formulae for
solutions of the equation
\[
F_1(x,u,{\mathop{u}\limits_{1}},\dots,{\mathop{u}\limits_{n}})=0.
\]

The approach to investigation of nonlocal symmetries of DEs
described above  allows treating  them as a basis for an algorithm
enabling construction of new solutions from one or more given
solutions.

In the following section we discuss in more detail application of
nonlocal symmetries to construction of explicit superposition
formulae.

\section{On symmetries and superposition principles\\ of nonlinear heat equations}

Nonlinear equations of the class
\[
u_t-\partial_x[h(u)+g(u)u_x]=f(x,u)
\]
were studied in a number of papers,
e.g.~\cite{dzamay&vorob'ev1994,clarkson&mansfield,pukhnachev,pukhnachev2,
sophocleous2004,sophocleous2005}.

Nonclassical (conditional) symmetries of nonlinear heat equations
of the general form
\[
u_t=u_{xx}+f(u)
\]
were completely described in~\cite{clarkson&mansfield}.
In~\cite{dzamay&vorob'ev1994} nonclassical partial symmetries of
equations from the family
\[
u_t=(g(u)u_x)_x+f(u)
\]
are constructed. These symmetries are  actually conditional
symmetries of the considered
equations~\cite{olver&rosenau1986,olver&rosenau1987,fushchych&serov}.

Nonlocal symmetries of some equations from the class
\[
u_t-\partial_x[\phi(u,u_x)]=0 %\qquad \mbox{or} \qquad u_t-\partial_x[C(u)u_x]=0
\]
were found by V.V.~Pukhnachev~\cite{pukhnachev,pukhnachev2} (in
particular, the conditions on the function $\phi$, under which the
transformation maps the equation into itself). The initial
equation can be transformed by the Lagrange transformation
\[
\xi=\int_0^xu(y,t)dy+\int_0^t\phi(u(0,s),u_x(0,s))ds,\qquad
\omega(\xi,t)=[u(x,t)]^{-1}
\]
into an equation from the same class
\[
\omega_t=[-\omega\phi(\omega^{-1},-\omega^{-3}\omega_{\xi})]_{\xi}.
\]

Further we will review dif\/ferent approaches to superposition
principles for solutions of nonlinear dif\/ferential equations.

One of examples of superposition principles for nonlinear
dif\/ferential equations is given by the Bianchi permutability
theorem for the sine-Gordon equation, which is adduced e.g.\
in~\cite{rogers&shadwick}. Permutability of auto-B\"acklund
transformations with dif\/ferent parameters allows constructing of
an inf\/inite family of soliton-like solutions for the sine-Gordon
equation by means of extension of two one-soliton solutions which
correspond to dif\/ferent values of parameters. The same method
was applied by H.D.~Wahlquist and F.B.~Estabrook for the
Korteveg--de Vries equation~\cite{wahlquist&estabrook}.

 The idea proposed by S.E.~Johnes and
W.F.~Ames~\cite{jones&ames} was developed
in~\cite{goard&broadbridge}. Two solutions $\overset{(1)}{u}$ and
$\overset{(2)}{u}$ of the nonlinear equation
\[
F(x,u,{\mathop{u}\limits_{1}},\dots,{\mathop{u}\limits_{n}})=0
\]
may be used by means of a procedure
\[
\Lambda(\overset{(1)}{u},\overset{(2)}{u})=\overset{(1)}{u}*\overset{(2)}{u}
\]
to construct a new solution $\overset{(3)}{u}$ of the same
equation:
\[
\Lambda:\quad(\overset{(1)}{u},\overset{(2)}{u})\to
\overset{(3)}{u}.
\]
(Here $*$ denotes a procedure combining these two solutions into a
new one.)

It was noted in~\cite{goard&broadbridge} that a superposition
principle is a symmetry of the system of two copies of the same
equation
\begin{gather}\label{syst.two eq-ns}
F(x,\overset{(1)}{u},\overset{(1)}{{\mathop{u}\limits_{1}}},\dots,
\overset{(1)}{{\mathop{u}\limits_{n}}})=0,\qquad
F(x,\overset{(2)}{u},\overset{(2)}{{\mathop{u}\limits_{1}}},\dots,
\overset{(2)}{{\mathop{u}\limits_{n}}})=0.
\end{gather}
They looked for symmetry operators of system~\eqref{syst.two
eq-ns} of the form
\begin{equation}\label{EqOpForSuperpos}
\Gamma=\eta(\overset{(1)}{u},\overset{(2)}{u})\partial_{\overset{(1)}{u}}
\end{equation}
whereas the representation symmetry operators of a single equation
in the classical approach are
\[
\Gamma=\eta(x,\overset{(1)}{u})\partial_{\overset{(1)}{u}}.
\]
The corresponding algorithm for generation of solutions is
implemented by means of calculating of the one-parameter
invariance group of system~(\ref{syst.two eq-ns})
\[
\overset{(3)}{u}={\overset{(1)}{u}}{}'=f(\overset{(1)}{u},\overset{(2)}{u};a),
\]
associated with the operator~\eqref{EqOpForSuperpos}. Here $a$ is
the group parameter. The algorithm actually represents a
realization of the solution superposition principle for the
equation under consideration.

The nonlinear heat equations of the class
\begin{gather}\label{general class non-lin.eq-ns}
 u_0+\partial_1(C_1(u)+C_2(u)u_1)=0
 \end{gather} remained
an object of interest for many
authors~\cite{moitsheki&broadbrige&edwards,galas,clarkson&mansfield}
for a long time. In particular, this  class of
equations~(\ref{general class non-lin.eq-ns}) includes the Burgers
equation
\begin{gather} \label{burgers eq-n}
u_0+uu_1-u_{11}=0
\end{gather}
and the nonlinear heat equation
\begin{gather}
\label{non-lin. heat eq-n} u_0-\partial_1(u^{-2}u_1)=0,
\end{gather}
which admit linearization by nonlocal transformation of variables.
It is well known that the equation~(\ref{burgers eq-n}) is
connected with linear heat equation
\begin{gather}\label{lin.heat eq-n}
\ v_0-v_{11}=0
\end{gather}
by the Cole--Hopf substitution
\begin{gather*}%\label{cole-hopf subst.}
u=-2(\ln v)_1.
\end{gather*}
The nonlocal transformation which provides linearization of
equation~(\ref{non-lin. heat eq-n}) was f\/irst found
in~\cite{storm} and then re-discovered in~\cite{bluman&kumei2}. It
can be presented in the
form~\cite{fushchych&serov&tychynin&amerov}
\begin{gather*}%\label{lin.transf.non-lin.heat eq-n}
u(x_0,x_1)=\frac1{v_1(y_0,y_1)}, \qquad x_1=v(y_0,y_1), \qquad
x_0=y_0.
\end{gather*}
Here $x=(x_0,x_1)$ and $y=(y_0,y_1)$ are tuples of old and new
independent variables correspondingly. This fact was utilized for
constructing of the nonlocal generating formulae and formulae of
nonlinear superposition principle for equation~(\ref{non-lin. heat
eq-n})
 in~\cite{fushchych&serov&tychynin&amerov} and for~(\ref{burgers
eq-n}) in this paper. These generating formulae are constructed
with symmetry operators of the linear heat equation~(\ref{lin.heat
eq-n}).

The invariance algebra of the linear heat equation~(\ref{lin.heat
eq-n}) was f\/irst calculated by S.~Lie and is adduced in the
standard texts e.g.\ by Ovsiannikov~\cite{ovsiannikov} and
Olver~\cite{olver1993}. A basis of this algebra consists of the
operators:
\begin{gather}  P_0=\partial_0,\qquad
P_1=\partial_1,\qquad I=v\partial_v, \qquad
 D=2y_0\partial_0+y_1\partial_1, \nonumber\\
 G=y_0\partial_1-\tfrac{1}{2}y_1v\partial_v,\qquad
 F=y_0^2\partial_0-y_0y_1-\tfrac{1}{4}(y_1^2+2y_0)v\partial_v,\qquad
 S_b=b(y_0,y_1)\partial_{v},\label{algebra lin. eq-n}
\end{gather}
where $b=b(y_0,y_1)$ is an arbitrary solution of~(\ref{lin.heat
eq-n}), i.e. $b_0=b_{11}$.

The operators $P_1$ and $G$ were applied
in~\cite{fushchych&serov&tychynin&amerov} to construction of
generating formulae for equation~(\ref{burgers eq-n}). The
operators $P_0$ and $P_1$ were used for the same purposes for
equation~(\ref{non-lin. heat eq-n}).

In the present paper new formulae for generating of solutions for
equations~(\ref{burgers eq-n}) and (\ref{non-lin. heat eq-n}) were
found. We also constructed new solutions of equation~(\ref{burgers
eq-n}) and~(\ref{non-lin. heat eq-n}) with the nonlocal
superposition formulae
\begin{gather}
\overset{(3)}{u}(x_0,x_1)=-2\partial_1\ln(\overset{(1)}{\tau}+\overset{(2)}{\tau}),\qquad
  -2\partial_1\ln(\overset{(k)}{\tau})=\overset{(k)}{u},\nonumber\\[1ex]
  -2\partial_0\ln\overset{(k)}{\tau}=\overset{(k)}{u_1}-\tfrac{1}{2}(\overset{(k)}{u})^2,
  k=1,2.\label{superpos.princ.burgers eq-n}
\end{gather}
for the Burgers equation
and~\cite{fushchych&serov&tychynin&amerov}
\begin{gather}
\overset{(3)}{u}(x_0,x_1)=\overset{(1)}{u^{-1}}
(x_0,\overset{(1)}{\tau})+\overset{(2)}{u^{-1}}(x_0,\overset{(2)}{\tau}),\qquad
  \overset{(1)}{u}(x_0,\overset{(1)}{\tau})d\overset{(1)}{\tau}
  =\overset{(2)}{u}(x_0,\overset{(2)}{\tau})d\overset{(2)}{\tau},\nonumber\\[1ex]
  x_1=\overset{(1)}{\tau}+\overset{(2)}{\tau},\qquad
  \overset{(k)}{\tau_0}=\overset{(k)}{{\tau_1}^{-2}}
  \overset{(k)}{\tau_{11}}\overset{(k)}{u^{-2}}(x_0,\overset{(k)}{\tau}),\quad
  k=1,2.\label{superpos.princ.non-lin. heat eq-n}
\end{gather}for
equation~(\ref{non-lin. heat eq-n}) . Here $\tau$'s denote the
functional parameters to be excluded from these formulae.

\section{Formulae for generating of solutions}

We can construct new formulae for generating of solutions of
linearizable equations by means of combining symmetry operator
formulae for generating of solutions of linear equations and
lineari\-zing transformations.

Namely, let $L$ be a linear dif\/ferential equation, $F$ a
nonlinear dif\/ferential equation and $\tau$ a~nonlocal
transformation which linearizes the equation~$F$ to the
equation~$L$. If $L$ admits a Lie symmetry operator $X$  being a
vector f\/ield on the space of independent and dependent variables
then the corresponding dif\/ferential operator $Q$ acting in the
space of unknown functions of~$L$ maps any solution
$\overset{(1)}{v}$ of~$L$ to the other solution
\begin{gather} \label{general nonloc. sym. lin. eq-n}
\overset{(2)}{v}=Q\overset{(1)}{v}.
\end{gather}
Then the nonlocal mapping~$\tau$ allows us to obtain a new
solution $\overset{(2)}{u}=\tau^{-1}\overset{(2)}{v}$ of~$F$ from
its known solution~$\overset{(1)}{u}=\tau^{-1}\overset{(1)}{v}$.
Here $\tau^{-1}$ is treated in a certain way. Therefore, in the
case of direct utilization of the transformation~$\tau$ for
generating of new solutions of~$F$ we have to make several steps
$\overset{(1)}{u}\to \overset{(1)}{v}\to \overset{(2)}{v}\to
\overset{(2)}{u}$. At the same time, it is possible to derive a
formula for generating of new solutions of~$F$ without involving
solutions of linear equations. This way is preferable since it
allows to avoid cumbersome calculations and to produce solutions
in one step.

Thus, combining the Cole--Hopf substitution linearizing the
Burgers equation~(\ref{burgers eq-n}) to the linear heat
equation~\eqref{lin.heat eq-n} and the action of the Lie symmetry
operator~$P_0=\partial_0$ on solutions of~\eqref{lin.heat eq-n},
we obtained  the following result.

\begin{theorem}
The generating formula for equation~\eqref{burgers eq-n}, which is
associated with the Lie symmetry operator~$P_0$ of~\eqref{lin.heat
eq-n}, has the form
\begin{gather} \label{gen. sol. form. burgers 1}
  \overset{(2)}{u}=\overset{(1)}{u}
  +\frac{\overset{(1)}{u_0}}{-\frac{1}{2}\overset{(1)}{u_1}+\frac{1}{4}(\overset{(1)}{u})^2}.
  \end{gather}
\end{theorem}

Applying the formula~(\ref{gen. sol. form. burgers 1})
iteratively, we construct chains of solutions of the Burgers
equation~(\ref{burgers eq-n}):
\begin{gather*}
 1. \ \ \frac{2}{e^{-x_0-x_1}-1} \to  -2 \to  \cdots\to  -2 \to  \cdots;\\
 2. \ \ -2\left[1+\frac{1}{x_1+2x_0+k}\right] \to  -2\left[1+\frac{1}{x_1+2x_0+k+2}\right] \\
 \phantom{2. \ \ }{} \to
-2\left[1+\frac{1}{x_1+2x_0+k+4}\right]  \to  \cdots,
\end{gather*}
here $k$ is an arbitrary constant;
\begin{gather*}
 3. \ \ \frac{x_1}{x_0} \to  \frac{x_1\left(6x_0-x_1^2\right)}{x_0\left(2x_0-x_1^2\right)}
 \to  \frac{x_1\left(60x_0^2-20x_0x_1^2+x_1^4\right)}{x_0\left(12x_0^2-12x_0x_1^2+x_1^4\right)} \\
 \phantom{ 3. \ \ }{}
 \to
 \frac{x_1\left(840x_0^3-420x_0^2x_1^2+42x_0x_1^4-x_1^6\right)}{x_0\left(120x_0^3-180x_0^2x_1^2+30x_0x_1^4-x_1^6\right)}
 \to  \cdots;\\
 4. \ \ -1-2\tanh (x_0+x_1)\! \to \! -\frac{13+14\tanh(x_0+x_1)}{5+4\tanh(x_0+x_1)}
\! \to \! -\frac{121 + 122\tanh(x_0+x_1)}{41+40\tanh(x_0+x_1)} \to
 \cdots.\!\!
 \end{gather*}

Consider the Lie symmetry operator~$P_0+P_1=\partial_0+\partial_1$
of~\eqref{lin.heat eq-n} and the corresponding operator formulae
for generating of solutions of the linear heat equation
\[ %\label{nonloc.sym.1 lin. eq-n}
   \overset{(2)}{v}=(\partial_0+\partial_1)\overset{(1)}{v}.
\]

 \begin{theorem}
The generating formula for equation~\eqref{burgers eq-n}, which is
associated with the Lie symmetry operator~$P_0+P_1$
of~\eqref{lin.heat eq-n}, has the form
\begin{gather} \label{gen. sol. form. burgers 2}
\overset{(2)}{u}=\overset{(1)}{u}-2\frac{\overset{(1)}{u_0}
+\overset{(1)}{u_1}}{\overset{(1)}{u_1}-\frac{1}{2}(\overset{(1)}{u})^2+\overset{(1)}{u}}.
\end{gather}
\end{theorem}

Similarly to the previous case, chains of solutions of
equation~(\ref{burgers eq-n}) can be obtained by
formula~(\ref{gen. sol. form. burgers 2}):
\begin{gather*}
1. \ \  C \to  C \to  \cdots;\\
2. \ \  \frac{2}{e^{-x_0-x_1}-1} \to  -2 \to -2
\to  \cdots;\\
3. \ \ -2\left[1+\frac{1}{x_1+2x_0+k}\right]\! \to
-2\left[1+\frac{2}{2x_1+4x_0+2k+3}\right] \!
\to  -2\left[1+\frac{1}{x_1+2x_0+k+3}\right]\!\\
\phantom{3. \ \ }{} \to
-2\left[1+\frac{2}{2x_1+4x_0+2k+9}\right]\!
\to  \cdots;\\
4. \ \ \frac{x_1}{x_0} \to
-\frac{-6x_0x_1+x_1^3-2x_1^2x_0+4x_0}{x_0(2x_0-x_1^2+2x_1x_0)}
 \\
 \phantom{4. \ \ }{} \to
\frac{24x_1x_0^3+48x_0^3-4x_1^3x_0^2-48x_1^2x_0^2
-60x_1x_0^2+4x_1^4x_0+20x_1^3x_0-x_1^5}{x_0(8x_0^3-4x_1^2x_0^2-24x_1x_0^2-12x_0^2+4x_1^3x_0+12x_1^2x_0-x_1^4)}
\to  \cdots;\\
5. \ \ -1-2\tanh(x_0+x_1) \! \to \! -\frac{23+22\tanh
(x_0+x_1)}{7+8\tanh (x_0+x_1)}\! \to \! -\frac{337+338\tanh
(x_0+x_1)}{113+112\tanh(x_0+x_1)} \! \to \cdots.
\end{gather*}

We also combine action of the Lie symmetry operator~$P_0+P_1$ on
solutions of~\eqref{lin.heat eq-n} with the linearizing
transformation of equation~\eqref{non-lin. heat eq-n} to the
linear heat equation~\eqref{lin.heat eq-n}.

\begin{theorem}
The generating formula for equation~\eqref{non-lin. heat eq-n},
which is associated with the Lie symmetry operator~$P_0+P_1$
of~\eqref{lin.heat eq-n}, has the form
\begin{gather}
\overset{(2)}{u}(x_0,x_1)=(\overset{(1)}{u})^5
\left[(\overset{(1)}{u_{\tau}})^2-\overset{(1)}{u_0}(\overset{(1)}{u})^3
-\overset{(1)}{u_{\tau}}(\overset{(1)}{u})^2\right]^{-1},\nonumber\\
  x_1=-\overset{(1)}{u_{\tau}}(\overset{(1)}{u})^{-3}+(\overset{(1)}{u})^{-1}.\label{gen. sol. form. non-lin. heat 1}
\end{gather}
\end{theorem}

After applying formula~(\ref{gen. sol. form. non-lin. heat 1}) to
the stationary solution
\[
\overset{(1)}{u}(x_0,x_1)=\frac{-1}{C_1x_1+C_2}
\]
of equation~(\ref{non-lin. heat eq-n}), we obtain the solution
\[
\overset{(2)}{u}(x_0,x_1)=\frac{-1}{C_1x_1}.
\]
Therefore, this solution is f\/ixed under action of~(\ref{gen.
sol. form. non-lin. heat 1}).

The next chosen operator from the Lie algebra~(\ref{algebra lin.
eq-n}) is  $D=2y_0\partial_0+y_1\partial_1$. It acts on solutions
of equation~\eqref{lin.heat eq-n} as
$\overset{(2)}{v}=2y_0\overset{(1)}{v_0}+y_1\overset{(1)}{v_1}$.

\begin{theorem}
The generating formula for equation~\eqref{non-lin. heat eq-n},
which is associated with the Lie symmetry operator~$D$
of~\eqref{burgers eq-n}, has the form
\begin{gather} \label{gen. sol. form. burgers 3}
\overset{(2)}{u}=\overset{(1)}{u}+\frac{2x_0\overset{(1)}{u_0}
+\overset{(1)}{u}+x_1\overset{(1)}{u_1}}{-x_0(\overset{(1)}{u_1}
-\frac{1}{2}(\overset{(1)}{u})^2)-\frac{x_1}{2}\overset{(1)}{u}}.
\end{gather}
\end{theorem}

The  following  chains of solutions of equation~(\ref{burgers
eq-n}) are obtained with the formula~(\ref{gen. sol. form. burgers
3}):
\begin{gather*}
1. \ \ C \to  C+\frac{1}{\frac{C}{2}x_0-\frac{1}{2}x_1} \to
C+\frac{4C^2x_0-4Cx_1+4}{C^3x_0^2-2C^2x_0x_1+4Cx_0+Cx_1^2-2x_1}
\to  \cdots;\\
2. \ \ \frac{2}{e^{-x_0-x_1}-1} \to
-2\left[1+\frac{1}{x_1+2x_0}\right] \to
-2\left[1+\frac{4x_0+2x_1+1}{4x_0^2+4x_0+4x_0x_1+x_1^2+x_1}\right]
\to  \cdots;\\
3. \ \ \frac{x_1}{x_0} \to  \frac{x_1}{x_0}
\to  \cdots;\\
4. \ \ -1-2\tanh (x_0+x_1) \to
-\frac{13x_0+5x_1+2+(14x_0+4x_1+4)\tanh
(x_0+x_1)}{5x_0+x_1+(4x_0+2x_1)\tanh (x_0+x_1)} \to  \cdots.
\end{gather*}

Now we construct new solutions of the equation~(\ref{burgers
eq-n}) by application of the superposition
principle~(\ref{superpos.princ.burgers eq-n}).

1. For the solutions $\overset{(1)}{u}=\frac{x_1}{x_0}$ and
$\overset{(2)}{u}=1+\frac{2}{x_0-x_1}$ we have
\[
\overset{(3)}{u}=\frac{C_1x_1e^{-\frac{x_1^2}{4x_0}}
-C_2e^{\frac{x_0}{4}-\frac{x_1}{2}}(x_0^{\frac{5}{2}}
-x_0^{\frac{3}{2}}x_1+2x_0^{\frac{3}{2}})}
{x_0(C_1e^{-\frac{x_1^2}{4x_0}}-C_2e^{\frac{x_0}{4}
-\frac{x_1}{2}}(x_0^{\frac{3}{2}}-x_0^{\frac{1}{2}}x_1))}.
\]

2. If $\overset{(1)}{u}=\frac{x_1}{x_0}$,
$\overset{(2)}{u}=-1-2\tanh (x_0+x_1)$ then
\[
\overset{(3)}{u}=-\frac{C_1x_1e^{-\frac{x_1^2}{4x_0}}
+iC_2e^{\frac{5x_0}{4}+\frac{x_1}{2}}x_0^{\frac{3}{2}}(2\sinh(x_0+x_1)
+\cosh (x_0+x_1))}{x_0(-C_1e^{-\frac{x_1^2}{4x_0}}+iC_2\cosh
(x_0+x_1)e^{\frac{5x_0}{4} +\frac{x_1}{2}}x_0^{\frac{1}{2}})}.
\]

3. The solutions $\overset{(1)}{u}=\frac{x_1}{x_0}$,
$\overset{(2)}{u}=\frac{2}{e^{-x_0-x_1}}$ generate the solution
\[
\overset{(3)}{u}=\frac{C_1x_1e^{-\frac{x_1^2}{4x_0}}-2C_2e^{x_0+x_1}x_0^{\frac{3}{2}}}
{x_0(C_1e^{-\frac{x_1^2}{4x_0}}-C_2x_0^{\frac{1}{2}}+C_2x_0^{\frac{1}{2}}e^{x_0+x_1})}.
\]

In a similar way, we construct new solutions of
equation~(\ref{non-lin. heat eq-n}) with superposition
principle~(\ref{superpos.princ.non-lin. heat eq-n}). Thus, from
the stationary solutions  $\overset{(1)}{u}=-\frac{1}{x_1}$ and
$\overset{(2)}{u}=-\frac{1}{2x_1}$ we obtain the solution
\[
\overset{(3)}{u}=\frac{2C_1e^{2x_0}}{-1-4C_1x_1e^{2x_0}+\sqrt{1+4C_1x_1e^{2x_0}}}.
\]
With other two stationary solutions
$\overset{(1)}{u}=\frac{1}{x_1}$,
$\overset{(2)}{u}=-\frac{1}{x_1+C}$ we get
\[
\overset{(3)}{u}=\left[-x_1+2e^{\left(LambertW\left(-\frac{1}{2}(x_1+C)e^{x_0-\frac{C_1}{2}}\right)
-x_0+\frac{C_1}{2}\right)}-C\right]^{-1}.
\]
Note that the constructed solutions are not stationary.

All obtained solutions can be extended to parametric families of
solutions with Lie symmetry transformations or by using other
known formulae of generating of solutions. For instance, for any
real value $r$ and any solution $v$ of the Burgers
equation~(\ref{burgers eq-n})
\[
\overset{(2)}{v}=-2\frac{\overset{(1)}{v_x}}{\overset{(1)}{v}+r}
\]
is a solution of the same Burgers equation.

\section[Classical and nonlocal symmetries of equation ${\rm Slid}(u_{\mu\nu})=0$]{Classical and nonlocal symmetries of equation $\boldsymbol{{\rm Slid}(u_{\mu\nu})=0}$}
% of the equation

Consider the equation
\begin{gather}\label{slid eq-n}
u_{11}u_{22}-u^{2}_{12}-(u_{00}u_{22}-u_{02}^{2})-(u_{00}u_{11}-u_{01}^{2})=0.
\end{gather}
It is an essentially nonlinear second order scalar PDE in three
independent variables, which can be linearized to the
(1+2)-dimensional d'Alembert equation. Its left hand side is the
sum of algebraic complements to the diagonal elements of the
adjoint matrix to the Hesse matrix~$(u_{\mu\nu})$ with the metric
tensor~$(g_{\mu\nu})={\rm diag}(1,-1,-1)$ of the Minkowski
space~$\mathbb{R}^{1,2}$. This equation was f\/irst investigated
in~\cite{fushchych&tychynin}. One of the authors
of~\cite{fushchych&tychynin} (W.I.~Fushchych) suggested the
notation
\[
{\rm Slid}(u_{\mu\nu}):= g_{\mu\nu}A^{\mu\nu}=0.
\]
Here $A^{\mu\nu}$ is the algebraic complement to the element
$u_{\mu\nu}$ of the Hessian matrix. Hereafter the indices~$\mu$
and~$\nu$ run from 0 to 2, and we assume summation over repeated
indices. To the best of our knowledge, the Lie symmetries of
equation~(\ref{slid eq-n}) had not been studied. We search for the
maximal Lie invariance algebra~$\mathcal{A}$ of
equation~(\ref{slid eq-n}) by the classical
technique~\cite{olver1993,ovsiannikov}.

In view of the inf\/initesimal invariance criterion, any operator
\begin{gather*}%\label{infinitesimal oper.}
X=\xi^\mu(x,u)\partial_\mu+\eta(x,u)\partial_{u}, \qquad
x=(x_{0},x_{1},x_{2}),
\end{gather*}
from~$\mathcal{A}$ satisfies the equation
\begin{gather*}%\label{necess. sufficient cond.}
X^{(2)}({\rm Slid}(u_{\mu\nu}))\big|_{M}=0,
\end{gather*}
where $X^{(2)}$ is the standard second-order prolongation of~$X$,
$M$ is the manifold determined by equation~(\ref{slid eq-n}) in
the second-order jet space. This is a necessary and suf\/f\/icient
condition for inf\/initesimal invariance of equation~(\ref{slid
eq-n}). We conf\/ine to the manifold $M$ e.g. with the equality
\begin{gather*}%\label{transition to m}
u_{22}=\frac{u_{00}u_{11}-u_{01}^{2}-u_{02}^{2}+u^{2}_{12}}{u_{11}-u_{00}}.
\end{gather*}
and then split the obtained expression with respect to the
unconstrained derivatives $u_{\mu\nu}$, $(\mu,\nu)\not=(2,2)$, and
$u_\mu$. After solving the derived system of determining
equations, we f\/ind the general form of coef\/f\/icients of
operators from~$\mathcal{A}$:
\begin{gather*}
\xi^{0}=C_1x_0+C_2x_1+C_3x_2+C_4,\nonumber\\
\xi^{1}=C_2x_0+C_1x_1-C_5x_2+C_6,\nonumber\\
\xi^{2}=C_3x_0+C_5x_1+C_1x_2+C_7,\nonumber\\
\eta=C_8x_0+C_9x_1+C_{10}x_2+C_{11}u+C_{12}.%\label{coordinates oper.}
\end{gather*}
Therefore, the maximal Lie invariance algebra~$\mathcal{A}$ of
equation~(\ref{slid eq-n}) is 12-dimensional. A basis
of~$\mathcal{A}$ consists of the operators
%\begin{gather}\label{algebra Al.12}\
%Al_{12}=\langle
%P_{0},P_{0},P_{1},P_{2},P_{3},J_{01},J_{20},J_{12},I,D,Q_{0},Q_{1},Q_{2}\rangle:
%\end{gather}
\begin{gather}
P_0=\partial_0,\qquad P_1=\partial_1,\qquad P_2=\partial_2,\qquad
P_3=\partial_u, \qquad I=u\partial_u, \nonumber
\\
J_{01}=x_{0}\partial_1+x_{1}\partial_0, \qquad
J_{02}=x_{0}\partial_2+x_{2}\partial_0, \qquad
J_{12}=-x_{2}\partial_1+x_{1}\partial_2,\nonumber\\
D=x_{0}\partial_0+x_{1}\partial_1+x_{2}\partial_2,\qquad
Q_{0}=x_{0}\partial_u,\qquad Q_{1}=x_{1}\partial_u, \qquad
Q_{2}=x_{2}\partial_u . \label{operators Al.12}
\end{gather}

The algebra~$\mathcal{A}$ can be used for f\/inding of exact
solutions of equation~(\ref{slid eq-n}). Exhaustive investigation
of Lie reductions of equation~(\ref{slid eq-n}) includes the
following steps: 1) construction of optimal systems of one- and
two-dimensional subalgebras of the algebra~$\mathcal{A}$ and
corresponding sets of ansatzes of codimensions one and two; 2)
reduction of the initial equations with these ansatzes to partial
dif\/ferential equations in two independent variables or ordinary
dif\/ferential equations; 3) solving of the reduced equations.

Here we restrict ourselves to particular reductions with respect
to one-dimensional subalgebras. Below we list basis operators of
these subalgebras, corresponding ansatzes, reduced equations for
the new unknown function~$\varphi$ in two independent variables
and some exact solutions of the initial equation, which are found
in the following way.

\medskip

1) $P_1+Q_{0}$:
\begin{gather}%\label{p1+q0 operator}
u=x_{0}x_{1}+\varphi(x_{0},x_{2}), \qquad
\varphi_{02}^{2}-\varphi_{00}\varphi_{22}+1=0;\nonumber
\\
\label{p1+q0 solution}
u=x_{0}x_{1}+\tfrac{1}{2}{x_{2}^{2}}+\tfrac{1}{2}{x_{0}^{2}}+C_1x_{0}+C_2.
\end{gather}

2) $P_1+Q_{1}$:
\begin{gather*}%\label{p1+q1 operator}
u=\tfrac{1}{2}{x_{1}^{2}}+\varphi(x_{0},x_{2}), \qquad
\varphi_{02}^{2}
-\varphi_{00}\varphi_{22}+\varphi_{22}-\varphi_{00}=0;
\\
u=\tfrac{1}{2}{x_{1}^{2}}\mp\tfrac{1}{2}\sqrt{(x_{2}^{2}-x_{0}^{2})(x_{2}^{2}-x_{0}^{2}+4C_1)}\nonumber\\
\phantom{u=}{}\mp
C_1\ln\left(2C_1+x_{2}^{2}-x_{0}^{2}+\sqrt{(x_{2}^{2}-x_{0}^{2})(x_{2}^{2}
-x_{0}^{2}+4C_1)}\right)-\tfrac{1}{2}(x_{2}^{2}-x_{0}^{2})+C_2.%\label{p1+q1 solution}
\end{gather*}

3) $J_{02}$:
\begin{gather*}%\label{i02 operator}
u=\varphi(\omega,x_{1}), \qquad \omega= x_{0}^{2}-x_{2}^{2},
\qquad
\omega(\varphi_{\omega1}^{2}-\varphi_{\omega\omega}\varphi_{11})+2\omega\varphi_{\omega}\varphi_{\omega\omega}-
\varphi_{\omega}\varphi_{11}+\varphi_{\omega}^{2}=0; %\label{i02 red. eq-n}
\\
u=-\tfrac{1}{2}{x_{1}^{2}}\pm\tfrac{1}{2}\sqrt{(x_{0}^{2}-x_{2}^{2})(x_{0}^{2}-x_{2}^{2}+4C_1)}\\
\phantom{u=}{} \pm
C_1\ln\left(2C_1+x_{0}^{2}-x_{2}^{2}+\sqrt{(x_{0}^{2}-x_{2}^{2})(x_{0}^{2}-x_{2}^{2}+4C_1)}
\right)-\tfrac{1}{2}(x_{0}^{2}-x_{2}^{2})+C_2.
\end{gather*}

4) $J_{02}+J_{12}$:
\begin{gather}%\label{i02+i12 operator}
u=\varphi(\omega_1,\omega_2),\qquad \omega_{1}= x_{0}+x_{1},\qquad
\omega_{2}= x_{2}^{2}+x_{1}^{2}-x_{0}^{2},\nonumber
\\%\label{i02+i12 red. eq-n}
\omega_{1}^{2}(\varphi_{12}^{2}-\varphi_{11}\varphi_{22})+4\omega_{2}\varphi_{2}\varphi_{22}+
4\omega_{1}\varphi_{2}\varphi_{12}+3\varphi_{2}^{2}=0;\nonumber
\\
\label{i02+i12 solution}
u=C_1\frac{\sqrt{x_{2}^{2}+x_{1}^{2}-x_{0}^{2}}}{x_{0}+x_{1}},
\qquad
u=\frac{C_2(x_{2}^{2}+x_{1}^{2}-x_{0}^{2})+C_1}{x_{0}+x_{1}}.
\end{gather}

Another way of application of Lie symmetries to construction of
exact solutions is generation of new solutions from known ones by
symmetry transformations. We can reconstruct the Lie symmetry
group~$\mathcal{G}$ of equation~(\ref{slid eq-n}) from the
algebra~$\mathcal{A}$ via solving of a set of Cauchy problems. As
a result, we obtained that the group~$\mathcal{G}$ is generated by
the following transformations:

\medskip

translations with respect to~$x$ and~$u$:\quad
$x'_\mu=x_\mu+a_\mu,\ u'=u+a_3$,

\medskip

scale transformations with respect to~$x$ and~$u$:\quad
$x'_\mu=e^{a_4}x_\mu,\ u'=e^{a_5}u$,

\medskip

addition of a linear in~$x$ term to~$u$:\quad $x'_\mu=x_\mu,\
u'=u+b_\mu x_\mu$,

\medskip

rotations in the plane $(x_1,x_2)$:\\ \null\qquad $x'_0=x_0,\
x'_1=x_1\cos c_1-x_2\sin c_1,\ x'_2=x_1\sin c_1+x_2\cos c_1,$
$u'=u$,

\medskip

Lorentz rotations in the plane $(x_0,x_1)$:\\ \null\qquad
$x'_0=x_0\cosh c_2+x_1\sinh c_2,\ x'_1=x_0\sinh c_2+x_1\cosh c_2,\
x'_2=x_2,$ $u'=u$,

\medskip

Lorentz rotations in the plane $(x_0,x_2)$:\\ \null\qquad
$x'_0=x_0\cosh c_3+x_2\sinh c_3,\ x'_1=x_1,\ x'_2=x_0\sinh
c_3+x_2\cosh c_3,$ $u'=u$,

\medskip

\noindent where $a_0, \ldots, a_5$, $b_\mu$, $c_1$, $c_2$ and
$c_3$ are arbitrary constants.

\medskip

Parametrical generation of solutions~(\ref{p1+q0 solution})
and~(\ref{i02+i12 solution}) by means of symmetry transformations
gives, for example, the following new solutions:
\begin{gather*}
%\label{solution tau i02}
u=(x_{2}a+x_{0})x_{1}\sqrt{1-a^2}+\tfrac{1}{2}{(x_{0}a+x_{2})^{2}}+\tfrac{1}{2}{(x_{2}a+x_{0})^{2}}
+C_1(x_{2}a+x_{0})+C_2,
\\
%\label{solution tau d}
u=\frac{C_2(x_{2}^{2}+x_{1}^{2}-x_{0}^{2})+C_1}{x_{0}\cosh
c_3+x_2\sinh c_3+x_{1}}.
\end{gather*}
Obtained solutions with parameters can be used in construction of
further new solutions, for instance, by application of the
nonlinear superposition principle.

\section{Nonlocal linearization and the formula of superposition\\ of
solutions for equation~(\ref{slid eq-n})}

As established in~\cite{fushchych&tychynin}, the (1+2)-dimensional
nonlinear equation $ {\rm Slid}(u_{\mu\nu}) =0$ can be reduced to
a linear one by the Legendre transformation in the space of
variables $x=(x_0,x_1,x_2)$:
\begin{gather}\label{legendre tr-n}
\begin{array}{l}
u(x_{0},x_{1},x_{2})=y_{0}v_{0}+y_{1}v_{1}+y_{2}v_{2}-v,\\
x_{0}=v_{0},
\\x_{1}=v_{1},\\x_{2}=v_{2}, \\
  \end{array}\qquad
v=v(y_{0},y_{1},y_{2}).
\end{gather}
The Legendre transformation is prolonged to the f\/irst and second
order derivatives in the space of variables $x$ in the following
way:
\begin{gather*}
u_{0}=y_{0},\qquad u_{1}=y_{1},\qquad u_{2}=y_{2},\nonumber\\
u_{00}=\frac{v_{11}v_{22}-v_{12}^{2}}{\delta},\qquad u_{01}=
-\frac{v_{01}v_{22}-v_{20}v_{12}}{\delta},\qquad u_{11}=\frac{v_{00}v_{22}-v_{20}^{2}}{\delta},\nonumber\\
u_{02}=\frac{v_{10}v_{21}-v_{11}v_{20}}{\delta},\qquad u_{01}=
-\frac{v_{00}v_{21}-v_{20}v_{10}}{\delta},\qquad
u_{22}=\frac{v_{00}v_{11}-v_{10}^{v_{12}}2}{\delta},
%\label{prolongation legendre tr-n}
\end{gather*}
where $\delta=\det(v_{\mu\nu})\neq0.$

Substituting the obtained expressions for $v_{\mu\nu}$
in~(\ref{slid eq-n}), we get the linear d'Alembert equation
\begin{gather}\label{d'Alembert}
\Box u=u_{00}-u_{11}-u_{22}=0.
\end{gather}
Since the Legendre transformation is an involution, i.e.\ it
coincides with its inverse transformation, application of the
Legendre transformation to~(\ref{d'Alembert}) results in the
initial equation~(\ref{slid eq-n}).

In~\cite{fushchych&serov&tychynin&amerov} a nonlinear
superposition formula was derived for any nonlinear dif\/ferential
equation which can be reduced by the Legendre
transformation~(\ref{legendre tr-n}) to a linear homogeneous
equation. In the case of equation~(\ref{slid eq-n}) this formula
has the form:
\begin{gather}\label{superp. form.slid}
\overset{(3)}{u}(x)=\overset{(1)}{u}(\tau)+\overset{(2)}{u}(\theta),\qquad
\overset{(1)}{u_{\mu}}(\tau)=\overset{(2)}{u_{\mu}}(\theta),\qquad
x=\tau+\theta,
\end{gather}
where $x=(x_{0},x_{1},x_{2})$, the subscript~$\mu$ of a
function~$u$ denotes dif\/ferentiation with respect to the
$\mu$-th argument of~$u$. $\tau=(\tau_0,\tau_1,\tau_2)$ and
$\theta=(\theta_0,\theta_1,\theta_2)$ are tuples of
parameter-functions depending on~$x$.

We apply formula~\eqref{superp. form.slid} for construction of new
solutions of~(\ref{slid eq-n}) from pairs of its known solutions.
Let us choose the solutions
\begin{gather*}%\label{initial sol.1}
\overset{(1)}{u}=-\sqrt{-2C_1(x_{1}^{2}-(x_{0}+ax_{2})^{2}-2(x_{0}+ax_{2}))+2C_1}+C_2,\\
%\label{initial sol.2}
\overset{(2)}{u}=-2\sqrt{-2C_1((ax_{0}+ax_{1}+x_{2})^{2}+(ax_{1}+ax_{2}+x_{0})^{2})\vphantom{\big|}}+C_2.
\end{gather*}
They satisfy the condition $\det (u_{\mu\nu})\neq 0$. Therefore,
it is possible to use formula~(\ref{superp. form.slid}) for
f\/inding of a new solution of~(\ref{slid eq-n}). We re-denote the
arguments of the f\/irst and second solutions as $\tau$
and~$\theta$ respectively. Replacing $\tau_\mu$ by the expression
$x_{\mu}-\theta_\mu$, we obtain the solution
\begin{gather*}%\label{result. sol.}
\overset{(3)}{u}=-\sqrt{-2C_1(x_1-\theta_1)^2+2C_1(a(x_2-\theta_2)+x_0-\theta_0)^2-(a(x_{2}-\theta_{2})+x_{0}-\theta_{0})+2C_1\vphantom{\big|}}\\
\phantom{\overset{(3)}{u}=}{}
-2\sqrt{2C_1((a\theta_{0}+a\theta_{1}+\theta_{2})^{2}+(a\theta_{1}+a\theta_{2}+\theta_{0})^{2})\vphantom{\big|}}+C_2.
\end{gather*}
The functional parameters  $\theta_{0}$, $\theta_{1}$ and
$\theta_{2}$ are found from the system
\begin{gather*}%\label{param. system}
R_1(-(a(x_{2}-\theta_{2})+x_{0}-\theta_{0})-1)
=R_2(\theta_{0}a^{2}+\theta_{1}a^{2}+2\theta_{2}a+\theta_{1}a+\theta_{0}),\\
R_1(x_{1}-\theta_{1})=R_2(\theta_{0}a^{2}+2\theta_{1}a^{2}+\theta_{2}a^{2}+\theta_{2}a+\theta_{0}a),\\
R_1(-(a(x_{2}-\theta_{2})+x_{0}-\theta_{0})-1)=R_2(\theta_{0}a^{2}+\theta_{1}a^{2}+2\theta_{2}a+\theta_{1}a+\theta_{0}),
\end{gather*}
where
\begin{gather*}%\label{defin. r1, r2}
R_1=\bigl(-2C_1(x_{1}-\theta_{1})^{2}+2C_1(a(x_{2}-\theta_{2})+x_{0}-\theta_{0})^{2}
-(a(x_{2}-\theta_{2})+x_{0}-\theta_{0})+2C_1\bigr)^{-\frac{1}{2}},\\
R_2=2\bigl(2C_1((a\theta_{0}+a\theta_{1}+\theta_{2})^{2}+(a\theta_{1}+a\theta_{2}+\theta_{0})^{2})\bigr)^{\frac{1}{2}}.
\end{gather*}

\section{Conclusion}

In this work we continued investigation of nonlocal symmetries of
PDEs by means of nonlocal transformations of variables.
Application of nonlocal symmetries to construction of explicit
superposition formulae and formulae of generation of solutions was
discussed. We obtained new formulae for generation of solutions
for the Burgers equation and the nonlinear heat equation. New
superposition principles were constructed for them and then used
for obtaining of new solutions. The formula of nonlinear
superposition of solutions for the multidimensional equation
\[
{\rm Slid}(u_{\mu\nu})
=u_{11}u_{22}-u^{2}_{12}-(u_{00}u_{22}-u_{02}^{2})-(u_{00}u_{11}-u_{01}^{2})=0
\]
was applied to construction of a new solution from two known ones.
Such algorithms of gene\-ra\-ting of new solutions represent new
nonlocal symmetries of nonlinear equations under investigation.

All obtained solutions can be extended to parametric families by
means of Lie symmetry transformations or by using other formulae
for generation of solutions.

\subsection*{Acknowledgments} The authors would like to thank the referees
for helpful suggestions and comments.

\pdfbookmark[1]{References}{ref}

\LastPageEnding
\end{document}